%% file: main.tex
\documentclass[conference]{IEEEtran}
\IEEEoverridecommandlockouts
\usepackage{cite}
\usepackage{amsmath,amssymb,amsfonts}
\usepackage{algorithmic}
\usepackage{graphicx}
\usepackage{textcomp}
\usepackage{xcolor}
\usepackage{booktabs}   
\usepackage{todonotes}  
\usepackage{enumitem}   
\usepackage{hyperref}   
\usepackage{soul}
\usepackage[stretch=10]{microtype} 

\usepackage[utf8]{inputenc}
\usepackage[T1]{fontenc}

\renewcommand*\descriptionlabel[1]{\hspace\labelsep
  \normalfont\bfseries{#1}}

\newcommand{\rsds}{\textsc{Rsds}}
\newcommand{\dask}{\textsc{Dask}}
\newcommand{\daskdistributed}{Dask/distributed}

\newcommand{\university}{IT4Innovations, VSB – Technical University of Ostrava}

\begin{document}

\title{Runtime vs Scheduler: Analyzing Dask's Overheads
\thanks{This work was supported by The Ministry of Education, Youth and Sports 
from the National Programme of Sustainability (NPS II) project “IT4Innovations 
excellence in science - LQ1602” and by The Ministry of Education, Youth and 
Sports from the Large Infrastructures for Research, Experimental Development, 
and Innovations project “e-INFRA CZ – LM2018140”. This work was also partially 
supported by the SGC grant No. SP2020/167 
"Extension of HPC platforms for executing scientific pipelines 2", VŠB - 
Technical University of Ostrava, Czech Republic.}
}

\author{
\IEEEauthorblockN{Stanislav Böhm}
    \IEEEauthorblockA{\textit{\university}\\stanislav.bohm@vsb.cz}
\and \IEEEauthorblockN{Jakub Beránek}
    \IEEEauthorblockA{\textit{\university}\\jakub.beranek@vsb.cz}
}

\maketitle


\begin{abstract}
Dask is a distributed task framework which is commonly used by data scientists 
to parallelize Python code on computing clusters with little programming 
effort. It uses a sophisticated work-stealing scheduler which has been 
hand-tuned to execute task graphs as efficiently as possible. But is scheduler 
optimization a worthwhile effort for Dask? Our paper shows on many real world 
task graphs that even a completely random scheduler is surprisingly competitive 
with its built-in scheduler and that the main bottleneck of Dask lies in its 
runtime overhead. We develop a drop-in replacement for the Dask central 
server written in Rust which is backwards compatible with existing Dask 
programs. Thanks to its efficient runtime, our server implementation is able to 
scale up to larger clusters than Dask and consistently outperforms it on a 
variety of task graphs, despite the fact that it uses a simpler scheduling 
algorithm.
\end{abstract}


\section{Introduction}
\label{sec:introduction}
\input{introduction}

\section{Related work}
\label{sec:relatedwork}
\input{related_work}

\section{Dask}
\label{sec:dask}
\input{dask}

\section{Reimplementing the Dask server}
\label{sec:rsds}
\input{rsds}

\section{Benchmarks}
\label{sec:benchmarks}
\input{benchmarks}

\section{Experiment evaluation}
\label{sec:evaluation}
\input{evaluation}

\section{Conclusion}
\label{sec:conclusion}
\input{conclusion}

\bibliographystyle{IEEEtran}
\bibliography{IEEEabrv,references}

\end{document}

%% file: introduction.tex
Distributed task frameworks are commonly used to scale programs to 
multiple nodes using little programming effort. While traditional 
HPC distributed paradigms like MPI promote fixed-size clusters with little 
resilience and low-level communication APIs designed for maximum performance, 
modern task frameworks like \dask{}~\cite{rocklin2015dask}, Ray~\cite{ray} or 
Spark~\cite{spark} support elastic clusters and provide high-level programming 
interfaces which abstract the communication aspect of the program away. 
They offer quick prototyping, even though they do not always provide the 
highest possible performance out-of-the-box. These frameworks are especially 
popular for distributing machine learning and data analysis programs with a few 
lines of code.

Even though each task framework offers its own set of APIs that enable writing 
distributed programs, they all eventually convert the input program into a 
\emph{task graph}. Vertices of this graph (called tasks) represent functions 
which operate on input data and generate output data. Arcs represent 
dependencies and data transfers between the tasks. This program representation 
is amenable to parallelization in a distributed environment. The job of a task 
framework is to decide on which computing nodes should the individual tasks be 
computed and manage data transfers between the nodes.

A crucial component of each task framework is the scheduler, which assigns 
tasks to nodes in order to minimize the total computation time. 
Finding the optimal task schedule is a well-known problem which is 
NP-hard even for very restricted formulations (e.g. even without network data 
transfers)~\cite{Ullman1975}. A number of heuristics 
have been proposed to tackle this problem, ranging from list-based scheduling 
to genetic algorithms 
~\cite{sih1993compile,wu1990hypertool,hlfet1974,kwok1999static,tang2010list,daoud2011hybrid,popa2018adapting}.
Many surveys and comparisons of these approaches were published 
in~\cite{kwok1998benchmarking,hagras2003static,wang2018list,estee}. 
Yet applying these algorithms to modern task frameworks is challenging, as they 
often make assumptions that do not hold in practice. For example they often 
assume that task durations are known in advance or that there is no network 
congestion. Task frameworks thus usually resolve to implementing their own 
scheduler with many heuristic decisions specific to their common use cases.

One example of such a framework is \dask{}, a popular Python library used to 
distribute and parallelize SQL-like table operations, machine learning 
workflows or even arbitrary Python functions with little programming effort. \dask{} uses a centralized architecture with a 
single central component -- the server -- which schedules tasks and manages 
computing nodes in a cluster. It uses a work-stealing scheduler which has been 
tuned extensively to support various task graphs. Yet it is unclear whether 
additional effort should be directed into improving the scheduler or if there 
is another bottleneck which should be prioritized.

We have designed a series of experiments that study the runtime overhead 
of \dask{} and the effect of the used scheduling algorithm on its performance. 
We also develop a backwards compatible replacement of the \dask{} server which 
aims to minimize its runtime overhead and thus scale to larger clusters. Our 
paper makes the following \emph{contributions}:
\begin{itemize}
    \item We demonstrate that even a very naïve scheduling algorithm, such as a 
    completely random scheduler, is in many common scenarios
    competitive with the sophisticated hand-tuned work-stealing scheduler used 
    by \dask{}.
    \item We develop \rsds{}, an open-source drop-in replacement for the 
    \dask{} server compatible with existing \dask{} programs that
    consistently outperforms the \dask{} server in a diverse benchmark set, 
    which we also openly released.
    \item We quantify and evaluate \dask{}'s task overhead using an 
    idealized worker implementation on various task graphs.
\end{itemize}

The structure of the paper is as follows. First we describe the architecture of 
\dask{} in section \ref{sec:dask} and the specifics of our server 
reimplementation in Rust\cite{Rust} in section \ref{sec:rsds}. We have designed 
several experiments to study the performance limits of \dask{} and compare its 
performance with \rsds{}. Our benchmarks are described 
in Section \ref{sec:benchmarks} and the evaluation is in Section 
\ref{sec:evaluation}. Lastly we conclude our findings in 
Section \ref{sec:conclusion}.

%% file: related_work.tex
A number of task framework benchmarks that included \dask{} have been 
published. The author of \dask{} has performed a series of \dask{} scaling 
microbenchmarks~\cite{daskscalingblog}. In~\cite{taskbench}, the authors 
benchmark several distributed task systems on a set of fundamental task graph 
shapes to compare their relative overhead and scaling properties. They find 
out that \dask{} stops scaling relatively quickly if the task granularity is 
smaller than ca. one hundred milliseconds, which we also confirm in our 
experiments. The performance of \dask{} vs 
Spark on a neuroimaging pipeline is compared in \cite{dasksparkcomparison}. Their results suggest that multi-threaded 
\dask{} workers have limited usability, which we also confirm in our 
evaluation. In \cite{sanzu}, a data science benchmark is performed that 
compares the performance of common Python, SQL, and R data analysis libraries 
on various task 
graphs. A task scheduler and a distributed runtime was developed 
in~\cite{hyperloom}. It compares its performance to \dask{} on one 
real-world and two synthetic task graphs. It is not backwards compatible 
with \dask{}.

None of the mentioned works examine the effect of the used scheduling algorithm 
in \dask{} nor do they implement minimal overhead baselines for the \dask{} 
server and worker.

%% file: dask.tex
\dask{} is a flexible Python library for parallel computation in Python. It 
offers APIs compatible with popular Python libraries like 
\texttt{pandas}\footnote{\url{https://pandas.pydata.org}} for table 
processing or \texttt{NumPy}\footnote{\url{http://numpy.org}} for 
n-dimensional array operations. Code that uses these interfaces is 
automatically transformed into a task graph, which is then executed in 
parallel. This enables almost transparent parallelization of sequentially 
looking Python code. Apart from these high-level interfaces, it is also 
possible to build the task graph manually from scratch.

\subsection{Task graph}
The core representation of a distributed program in \dask{} is the \emph{task 
graph}. It is a directed acyclic graph (DAG), where the vertices (tasks) are 
Python functions that should be computed and arcs are dependencies between 
tasks. An arc from a task $u$ to a task $v$ represents that task $v$ needs to use the output produced by task $u$ as one of its input parameters. Therefore $u$ has to be finished before $v$ can begin executing and the output of $u$ has to be transferred over the network if $v$ is scheduled to a different computing node than $u$.

\subsection{Architecture}
\dask{} supports multiple computing backends. In this work we focus 
solely on \daskdistributed{}\footnote{\url{https://distributed.dask.org}}, 
which computes task graphs on a distributed cluster. When we refer to \dask{}, 
we assume that it uses this distributed backend. Its architecture is composed 
of three main components: the \emph{client}, the \emph{server} and the 
\emph{worker}.

\emph{Client} is the user-facing API used to run distributed computations. The 
client code connects to a \dask{} cluster, submits task graphs to the server 
and gathers the results.

\emph{Server} is the central component of the \dask{} cluster which coordinates 
workers and handles requests from clients. It contains a scheduler which 
assigns tasks to workers to load balance the cluster. Each \dask{} cluster has 
a single server.

\emph{Worker} is a process which executes tasks (Python functions) 
submitted by the client. Server sends tasks (consisting of serialized Python 
functions and arguments) to individual workers to be executed. Workers 
communicate amongst themselves to exchange task outputs that are not available 
locally. Each worker is configured with a number of CPU cores that it is 
allowed to use. Workers process their tasks in parallel, but they never execute 
more than one task per available core at once.

\subsection{Programming interface}
\label{sec:daskapi}
\dask{} allows defining computational workflows in several ways. At the lowest 
level is the \emph{Futures} interface which can lazily build 
a task graph from Python functions. On top of Futures \dask{} offers high-level 
interfaces that mimic the API of popular Python libraries. Examples of these 
include \emph{Arrays}, a parallel version of the \texttt{NumPy} 
library for numerical computations on tensors or \emph{DataFrame}, 
a parallel version of the \texttt{Pandas} library for analysis of 
tabular data.

These interfaces use the concept of \emph{partitions}. It is a
parameter which controls the granularity of tasks created by \dask{}. For 
example, an operation on a large table might be partitioned into several tasks, 
each operating on a subset of rows. The selected amount of partitions specifies 
into how many parts will the table be divided. It is crucial to choose 
this parameter accurately, otherwise the task graph might not be parallelized 
effectively.

\subsection{Work-stealing scheduler}
\dask{} uses a sophisticated work-stealing scheduler containing many heuristics
that have been tuned for quite some time. Some of them are described in 
the \dask{} 
manual\footnote{\url{https://distributed.dask.org/en/latest/work-stealing.html}}.
The scheduler works in the following way: when a task becomes ready, 
i.e. all its dependencies are finished, it is immediately assigned to a 
worker according to a heuristic that tries to minimize an estimated start time 
of the task. The estimate is based on potential data transfers and the current 
occupancy of workers. When an imbalance occurs (some workers are 
underloaded/overloaded), the scheduler tries to steal tasks from overloaded 
nodes and 
distribute them to underloaded nodes. The scheduler also assigns priorities 
to tasks that are used by workers to decide what tasks should be computed 
first.

\subsection{Integrating schedulers into \dask{}}
Integrating existing task scheduling algorithms into \dask{} is difficult, 
because the work-stealing concept is integrated tightly inside \dask{}.
Integrating an existing algorithm would thus require making modifications both 
to \dask{} and to the algorithm via heuristic decisions which could introduce 
bias.

To avoid that, we have implemented possibly the simplest possible scheduling 
algorithm -- a random scheduler. A completely random scheduler does not need 
any additional heuristics and is therefore immune from any possible bias that 
we could introduce by our implementation. It is also so simple that it could be 
included in \dask{} without making large modifications to it. Our random 
scheduler eagerly assigns each task to a random worker using a uniform random 
distribution.

%% file: rsds.tex
To examine the runtime overhead of \dask{}, we needed a baseline for the server 
implementation with the lowest possible overhead. To get close to this 
goal, we have reimplemented the server in Rust. Rust has a minimal runtime and 
manual memory management, which by itself reduces the ubiquitous overhead of 
reference counting and indirection present in Python. It also has direct 
programming support for asynchronous I/O and it provides strong memory safety 
guarantees, hence it is well suited for writing distributed applications. We 
call our server reimplementation \rsds{}.

\rsds{} is compatible with existing \dask{} clients and workers and it can thus 
be used to run \dask{} programs without making any source code changes. Even 
though it is not feature complete, it supports a minimum set of \dask{} 
message types which are necessary to run the most common \dask{} workflows. 
\rsds{} is able to run a diverse set of unmodified \dask{} programs described 
in Section \ref{sec:benchmarks}. We provide \rsds{} as open-source software at 
\url{https://github.com/it4innovations/rsds}.

\subsection{Architecture}
\begin{figure}[tbph]
\centering
\includegraphics[width=0.8\linewidth]{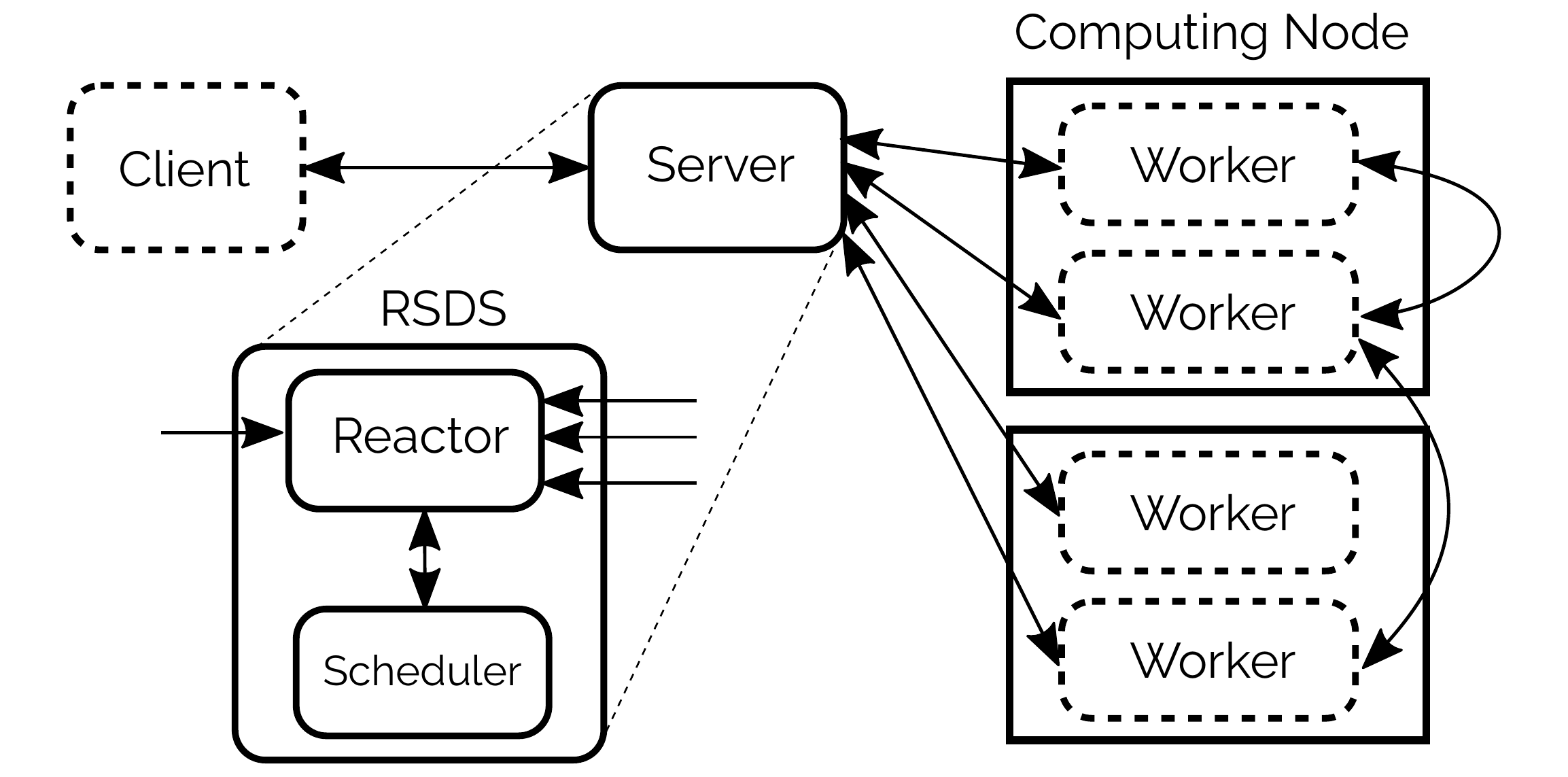}
\caption{Architecture of \rsds{} (Dask components are dashed)}
\label{fig:arch}
\end{figure}

The architecture of \rsds{} is shown in Figure~\ref{fig:arch}.
The main architectural difference between \rsds{} and \dask{} is the 
separation of the server into two parts: \emph{reactor} and 
\emph{scheduler}. The reactor manages worker and client connections, maintains 
bookkeeping information and translates scheduler assignments into \dask{} 
messages which are then sent to the workers.

The scheduler is a process which receives a task graph and outputs 
assignments of tasks to workers. It does not care about network connections, 
the \dask{} protocol messages or any other bookkeeping that is not relevant to 
scheduling. Since it is isolated, it can be swapped easily and therefore it is 
trivial to experiment with different schedulers in \rsds{}. Another benefit is 
that we can easily run the scheduler in a separate thread to enable concurrent 
scheduling and runtime management. This is possible because the scheduler does 
not share any data structures with the reactor. A disadvantage of this scheme 
is that the both the reactor and the scheduler need to build their own task 
graph, which increases memory usage, but we have not found this to be a problem 
in practice.

\subsection{\dask{} protocol modifications}



\dask{} uses a custom language-agnostic communication protocol serialized by
MessagePack~\footnote{\url{https://msgpack.org}}. \dask{} sometimes arbitratily 
fragments message structures into parts and then reassembles them during 
deserialization. This is difficult to perform in a statically typed language. 
We have modified it to keep the original message structure and replace values 
that must be fragmented with placeholders, which are later replaced with 
deserialized values. This avoids the need to dynamically change the message 
structure during deserialization.


Our change only modifies low-level message handling, it is thus fully 
transparent to the rest of the code and spans less than 100 modified lines of 
\dask{} source code. It has no effect on the functionality of clients, workers 
and the server. Our modified version of \dask{} is open-source and available online\footnote{\url{https://github.com/kobzol/distributed/tree/simplified-encoding}}.
All evaluations presented in this paper use the modification described above, 
even when \rsds{} is not used. We have benchmarked this modification and found that there are no performance differences in respect to the original \dask{} message encoding.

\subsection{Schedulers}
We have implemented two schedulers in \rsds{} -- a work-stealing scheduler and 
a random scheduler -- to test the runtime overhead difference between \dask{} 
and \rsds{} while using a similar set of schedulers.

Even though it was not possible to exactly replicate the work-stealing 
implementation used in \dask{}, our implementation is inspired by it. However,
it is also deliberately simple to avoid the need to perform extensive 
hand-tuning. Some of the heuristics used by \dask{} were changed, simplified, 
or dropped in our implementation. For example \rsds{} does not 
estimate average task durations and does not use any network speed estimates.

The \rsds{} work-stealing scheduler works as follows: when a task 
becomes ready (i.e. all its inputs are already finished), it is immediately 
assigned to a worker.
The scheduler chooses a worker where the task may be executed with minimal 
data transfer costs, while it deliberately ignores the load of the worker.
The load is ignored to speed up the decision in optimistic situations when 
there is enough tasks to keep the workers busy. When it is not the case, it is 
solved by balancing (described below).

For computing transfer costs, we use a heuristic that takes into account
inputs that are already present on the worker's node and also inputs that will 
be eventually present because they are in transit or they are depended upon by 
another task assigned to the same worker. Transfer cost is smaller for data 
transfers between workers residing on the same node.

When a new task is scheduled or when a task is finished, the scheduler checks 
if there are nodes that are under-loaded. In such case, balancing is performed 
and the scheduler reschedules tasks from workers with sufficient number of tasks
to under-loaded workers. During rescheduling, the scheduler simply passes the 
new scheduling state to the reactor, which performs all of the complex 
rescheduling logic. It tries to retract rescheduled tasks from their originally 
assigned workers. If retraction succeeds, the task is scheduled to the newly 
assigned worker. When the retraction fails, because a task is already running 
or has been finished, the scheduler is notified and it then initiates balancing 
again if necessary.

Our random scheduler mirrors the random scheduler implementation in 
\dask{} -- it assigns a random worker using a uniform random distribution to 
each task as soon as the task arrives to the server. It ignores any other 
scheduling mechanisms, such as task stealing, and does not maintain any task 
graph state.

\subsection{Zero worker}
\label{sec:zero-worker}
To quantify the runtime overhead of the \dask{} server, we have created an 
implementation of the \dask{} worker that we call \emph{zero worker}. Zero 
worker is a minimal implementation of the \dask{} worker process written 
in Rust. Its purpose is to simulate a worker with infinite computational speed, 
infinitely fast worker-to-worker transfers and zero additional overhead. It 
actually does not perform any real computation; when a task is assigned to a 
zero worker, it immediately returns a message that the task was finished. It 
also remembers a set of data-objects that would be placed on the worker in a 
normal computation. When a task requires a data object which is not in this 
list, the worker immediately sends a message to the server that the object was 
placed on it -- this simulates an infinitely fast download of data between 
workers. No actual worker-to-worker communication is performed in such case.

Zero workers respond to every data fetch request with a small mocked constant 
data object. Such requests come from the server when a client asks for a data 
object, usually at the end of the computation. Since there is no 
worker-to-worker communication, fetch requests never come from other workers.

We use the zero worker in our experiments only when explicitly stated; 
otherwise, we always use the original \dask{} worker implementation.

%% file: benchmarks.tex
This section covers task graphs benchmarked in our experiments. We have 
prepared a diverse set of benchmarks that span from simple map-reduce 
aggregations to text processing workloads and table queries. The properties of the task graphs used in our experiments along with 
the \dask{} API that was used to create them are summarized in Table  
\ref{tab:graph_properties}. Most of the task graphs are heavily inspired by 
programs from the \dask{} Examples 
repository\footnote{\url{https://examples.dask.org}}. Our 
benchmark dataset is available in the \rsds{} repository. A short summary of the individual benchmarks is provided below. 

\noindent\textbf{merge-n} creates $n$ independent trivial tasks that are merged at the end. This benchmark is designed to stress the scheduler and the server.

\noindent\textbf{merge\_slow-n-t} is similar to \texttt{merge-n}, but with longer, $t$ second tasks.

\noindent\textbf{tree-n} performs a tree reduction of $2^n$ numbers using a 
binary tree with height $n-1$.

\noindent\textbf{xarray-n} calculates aggregations (mean, sum) on a 
three-dimensional grid of air temperatures~\cite{airdataset}, $n$ specifies 
size of grid partitions.

\noindent\textbf{bag-n-p} works with a dataset of $n$ records in $p$ partitions. It performs a cartesian product, filtering and aggregations.

\noindent\textbf{numpy-n-p} transposes and aggregates a two-dimensional 
distributed NumPy array using the \emph{Arrays} interface. The array has size 
$(n,n)$ and it is split into partitions of size $(n/p,n/p)$.

\noindent\textbf{groupby-d-f-p} works with a table with $d$ days of records, each 
record is $f$ time units apart, records are partitioned by $p$ time units. It 
performs a groupby operation with an aggregation.

\noindent\textbf{join-d-f-p} uses the same table, but performs a self-join.

\noindent\textbf{vectorizer-n-p} uses  
Wordbatch\footnote{\url{https://github.com/anttttti/Wordbatch}}, a text 
processing 
library, to compute hashed features 
of $n$ reviews from a TripAdvisor dataset~\cite{wordbatcharticle} split into 
$p$ partitions.

\noindent\textbf{wordbag-n-p} uses the same dataset, but computes a full text processing pipeline with text normalization, spelling correction, word counting and feature extraction.

\setlength{\tabcolsep}{5pt}
\begin{table}
    \caption{Task graph properties}
    \centering
    \label{tab:graph_properties}
\begin{tabular}{l|rrrrrc}
    \toprule
    \textbf{Task graph} & \textbf{\#T} & \textbf{\#I} & \textbf{S} & 
    \textbf{AD} & \textbf{LP} & \textbf{API} \\
    \midrule
merge-10K & 10001 & 10000 & 0.027 & 0.006 & 1 & F \\
merge-15K & 15001 & 15000 & 0.027 & 0.006 & 1 & F \\
merge-20K & 20001 & 20000 & 0.027 & 0.006 & 1 & F \\
merge-25K & 25001 & 25000 & 0.027 & 0.006 & 1 & F \\
merge-30K & 30001 & 30000 & 0.027 & 0.006 & 1 & F \\
merge-50K & 50001 & 50000 & 0.027 & 0.006 & 1 & F \\
merge-100K & 100001 & 100000 & 0.027 & 0.006 & 1 & F \\
merge\_slow-5K-0.1 & 5001 & 5000 & 0.023 & 100 & 1 & F \\
merge\_slow-20K-0.1 & 20001 & 20000 & 0.023 & 100 & 1 & F \\
tree-15 & 32767 & 32766 & 0.027 & 0.007 & 14 & F \\
xarray-25 & 552 & 862 & 55.7 & 3.1 & 10 & X \\
xarray-5 & 9258 & 14976 & 3.3 & 0.4 & 10 & X \\
bag-25K-10 & 236 & 415 & 292 & 1233 & 6 & B \\
bag-25K-100 & 21631 & 41430 & 3.2 & 13.9 & 8 & B \\
bag-25K-200 & 86116 & 165715 & 0.8 & 3.6 & 9 & B \\
bag-25K-50 & 5458 & 10357 & 12.6 & 54.9 & 7 & B \\
bag-50K-50 & 5458 & 10357 & 25.2 & 214 & 7 & B \\
numpy-50K-10 & 209 & 228 & 70108 & 169 & 7 & A \\
numpy-50K-100 & 19334 & 21783 & 760 & 2.6 & 10 & A \\
numpy-50K-200 & 77067 & 86966 & 191 & 0.9 & 11 & A \\
numpy-50K-50 & 4892 & 5491 & 2999 & 8.3 & 9 & A \\
groupby-2880-1S-16H & 22842 & 31481 & 1005 & 11.9 & 9 & D \\
groupby-2880-1S-8H & 45674 & 62953 & 503 & 7.7 & 9 & D \\
groupby-1440-1S-1H & 182682 & 251801 & 64.3 & 3.8 & 10 & D \\
groupby-1440-1S-8H & 22842 & 31481 & 503 & 7.7 & 9 & D \\
groupby-360-1S-1H & 45674 & 62953 & 64.3 & 3.8 & 9 & D \\
groupby-360-1S-8H & 5714 & 7873 & 503 & 8.0 & 8 & D \\
groupby-90-1S-1H & 11424 & 15743 & 64.3 & 3.9 & 8 & D \\
groupby-90-1S-8H & 1434 & 1973 & 501 & 7.7 & 7 & D \\
join-1-1S-1H & 673 & 1224 & 15.3 & 33.0 & 5 & D \\
join-1-1S-1T & 72001 & 125568 & 3.7 & 1.7 & 11 & D \\
join-1-2s-1H & 673 & 1224 & 9.3 & 9.8 & 5 & D \\
vectorizer-1M-300 & 301 & 0 & 10226 & 1504 & 0 & F \\
wordbag-100K-50 & 250 & 200 & 5136 & 301 & 2 & F \\
\bottomrule
    \end{tabular}\\
    \vspace{1mm}
	
    \#T = Number of tasks; \#I = Number of dependencies; \\
    S = Average task output size [KiB]; AD = Average task duration [ms]; \\
    LP = longest oriented path in the graph; \\
    D = DataFrame; B = Bag; A = Arrays; F = Futures; X = XArray
\end{table}

%% file: evaluation.tex
This section presents our evaluation setup and methodology and discusses  the 
results of several experiments that we have designed to test \dask{} schedulers 
and compare our \rsds{} server with the original \dask{} server implementation.

All experiments were performed on the Salomon 
supercomputer\footnote{\url{https://docs.it4i.cz/salomon/introduction}}. Each 
Salomon node has two sockets containing Intel Xeon E5-2680v3 with 12 cores 
clocked at 2.5 GHz (24 cores in total), 128 GiB of RAM clocked at 2133 MHz and 
no local disk. The interconnections between nodes use InfiniBand FDR56 with 7D 
enhanced hypercube topology.

In all of our experiments, we use a setting with 24 \dask{} workers per node, 
each using a single thread for task computations. We chose this setting because 
of the Global Interpreter Lock present in the standard CPython interpreter. 
Since our benchmarks are compute-bound and not I/O-bound, a single worker 
cannot effectively use more than a single thread. Not even the popular 
\texttt{NumPy} and \texttt{Pandas} libraries used in our benchmarks are 
multi-threaded by default, which is also why \dask{} provides direct API 
support for their parallelization. To confirm our decision, we have 
benchmarked a configuration using a single worker with 24 threads per each 
node. We have found that it provides no benefit in comparison to a 
single worker with only one thread in any of our benchmarks. These tests are 
not reported in the paper. The fact that some workflows do not benefit from 
multi-threaded \dask{} workers has been also observed 
in~\cite{dasksparkcomparison}.

For each of our experiments, we state the number of used worker nodes, these 
contain only the workers. We always use one additional node which runs both the 
client and the server. For our scaling experiments we use 1 to 63 worker
nodes (24-1512 \dask{} workers), for the rest of our experiments we use 
either 1 or 7 worker nodes (24 or 168 \dask{} workers). We have chosen these 
two cluster sizes to represent a small and a medium sized \dask{} cluster. The 
number of workers is fixed, they do not connect nor disconnect during the 
computation. The timeout for all benchmarks was set to 300 seconds.

We have run each benchmark configuration five times (except for the scaling 
benchmarks, which were executed two times to lower our computational budget) 
and averaged the result. We measure the duration between the initial task graph 
submission to the server and the processing of the final output task by the 
client. We call this duration \emph{makespan}. We reset the whole cluster 
between each benchmark execution.

We use the following abbreviations in figures with benchmark results: \emph{ws} 
marks the work-stealing scheduler and \emph{random} represents the random 
scheduler.

\subsection{Scheduler}
\label{sec:daskschedulerexperiment}
\begin{figure}
    \centering
    \includegraphics[width=0.5\textwidth]{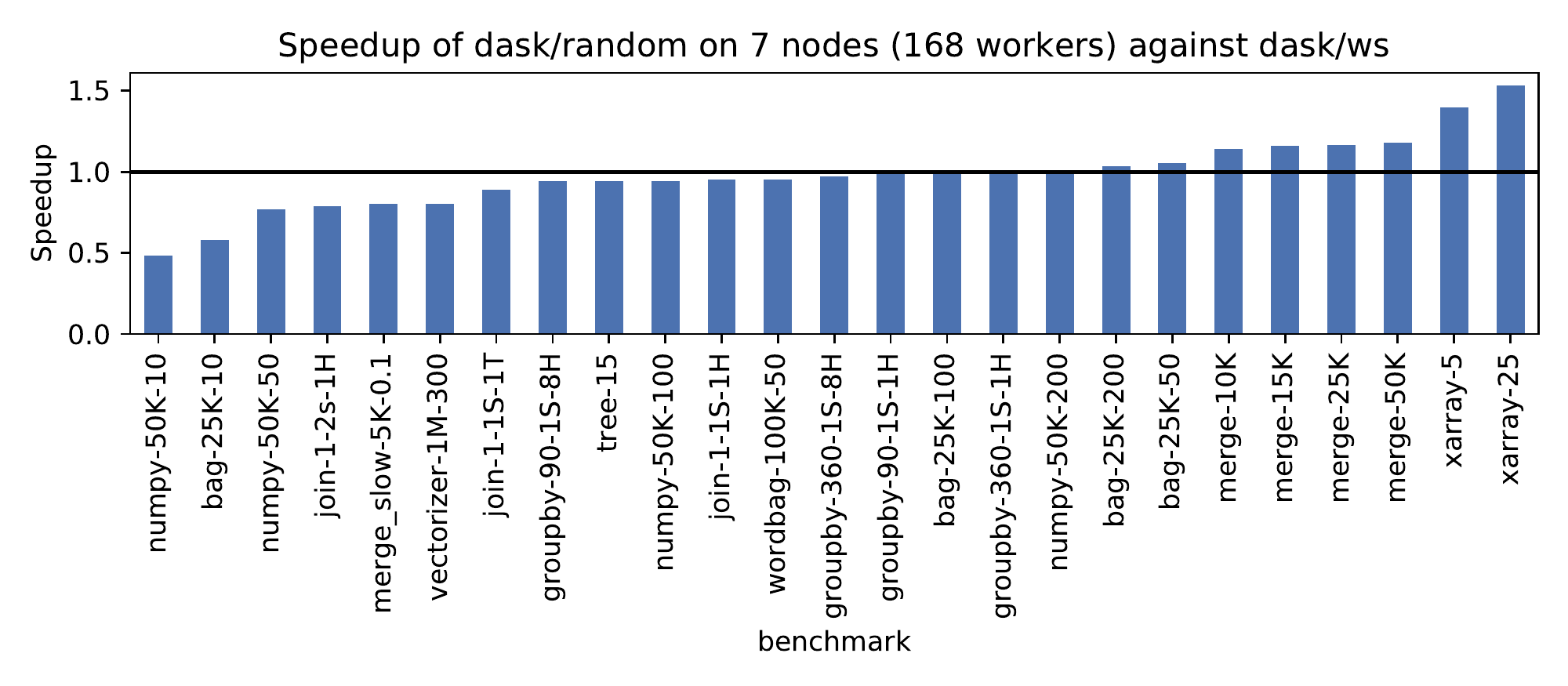}
    \caption{Speedup of \dask{}/random scheduler;
    \dask{}/ws is baseline.}
    \label{fig:daskrandom-all}
\end{figure}

Our first experiment compares the built-in work-stealing scheduler vs a 
random scheduler used in the \dask{} server. The overall results are shown in 
Figure~\ref{fig:daskrandom-all}. To our surprise, we have found out that the 
simple random scheduler fares relatively well on both small and medium sized 
clusters. At worst, it produces a twice longer makespan, but overall it is 
quite close to the performance of the work-stealing scheduler and in some cases 
it even outperforms it with a $1.4\times$ speedup. The geometric mean of 
speedup in comparison to the work-stealing scheduler over all tested task 
graphs can be observed in Table~\ref{tab:geom_mean_speedup}. It can be seen 
that the random scheduler gets closer to the performance of work-stealing if 
more workers are used.

There are two reasons for this. Firstly, with more workers, the work-stealing 
scheduler has to do more work to compute where should the tasks be computed and 
it also generates additional network traffic by sending task stealing messages. 
The random scheduler has a fixed computation cost per task independent of the 
worker count. It simply chooses a worker randomly and it sends no additional 
messages other than one assignment per task. The second reason is that our 
benchmark set is more computationally bounded rather than network bounded. 
Therefore, with more workers, if the scheduler wants to utilize all 
computational power of the cluster, network transfers are less avoidable. This 
decreases the chance that a random scheduler induces an unnecessary data 
transfer that could have been avoided by a smarter scheduler.

The results of this experiment suggest that in many scenarios, a complex 
scheduling algorithm is not needed and a random schedule is sufficient for 
\dask{}. This opens up the question whether the schedule is the main bottleneck 
which slows down the execution of \dask{} workflows. We try to answer it in the 
following experiments.

\subsection{Server}
\begin{figure}
    \centering
    \includegraphics[width=0.5\textwidth]{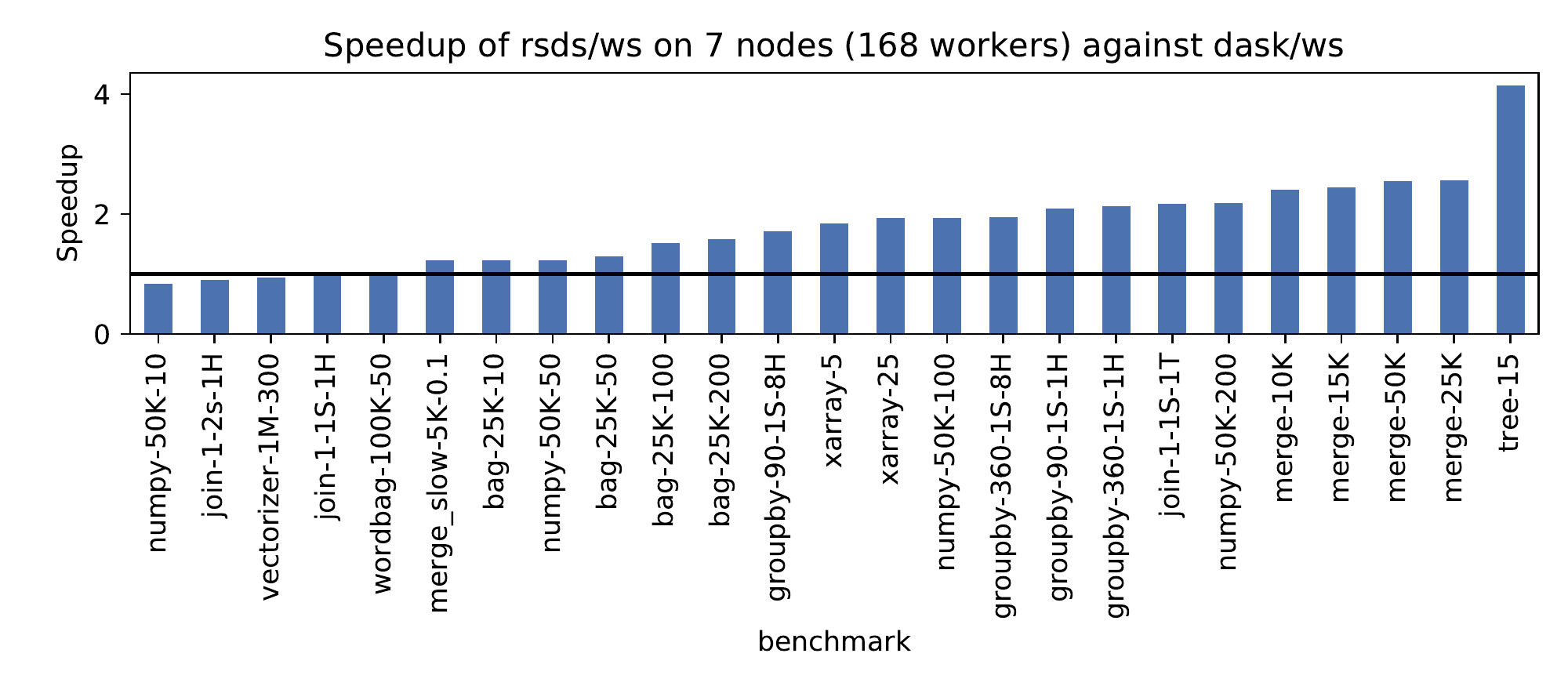}
    \caption{Speedup of \rsds{}/ws scheduler; \dask{}/ws is baseline.}
    \label{fig:ws-all}
\end{figure}

\begin{figure}
    \centering
    \includegraphics[width=0.5\textwidth]{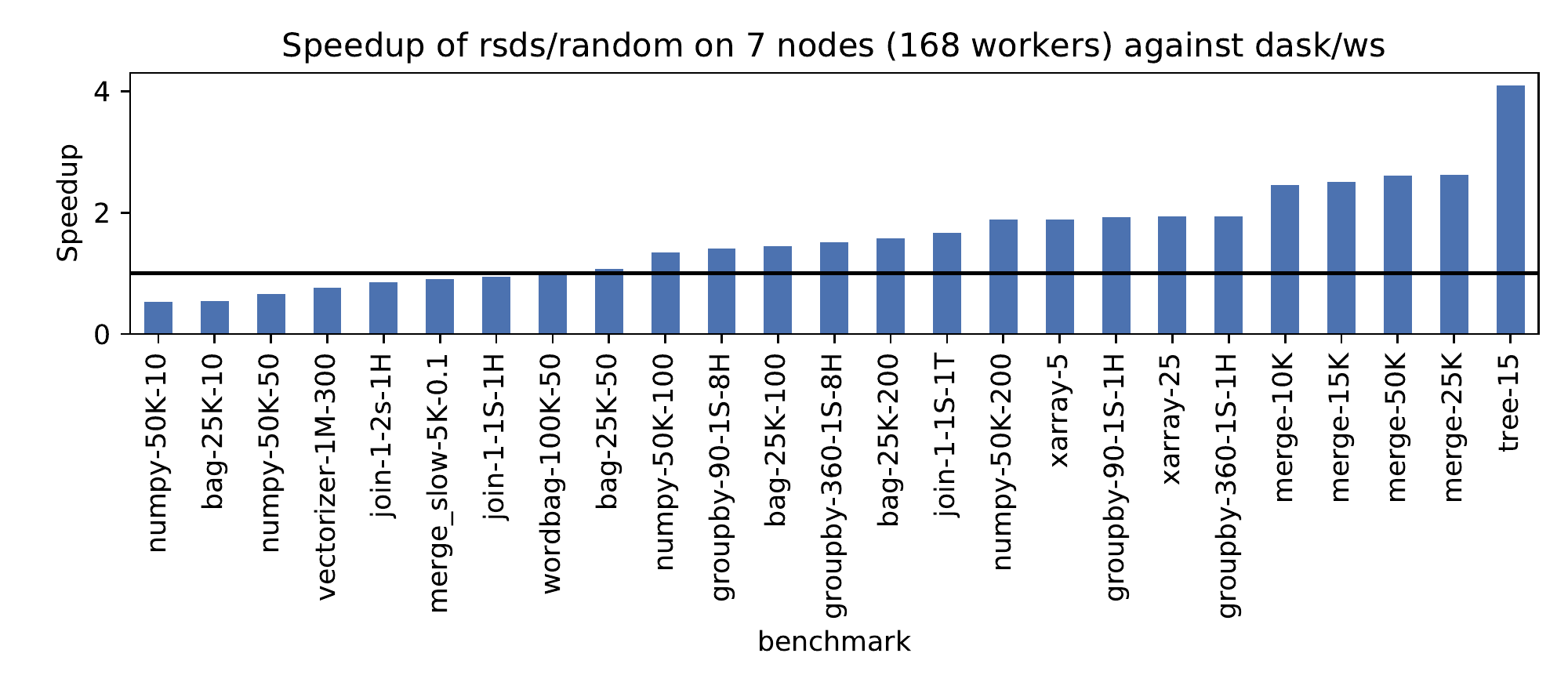}
    \caption{Speed up \rsds{}/random scheduler; 
    \dask{} work-stealing is baseline.}
    \label{fig:dask-ws-rsds-random-all}
\end{figure}

This experiment compares the efficiency of \rsds{} and the original \dask{} 
server with both the work-stealing and the random scheduler. Results for 
work-stealing schedulers are shown in Figure~\ref{fig:ws-all}. The 
data confirm our expectation that \dask{} has large inherent overhead and that 
reducing it helps to improve the makespan of executed task graphs. Even though 
\rsds{} uses a much simpler work-stealing scheduler, its more efficient runtime 
provides better performance in most cases and this effect becomes even more 
visible with larger clusters. Figure~\ref{fig:dask-ws-rsds-random-all} 
confirms that the speedup is caused by the reduced runtime overhead and not by 
the different work-stealing implementation. It shows the speedup of \rsds{} 
using a random scheduler over \dask{} with the work-stealing scheduler. 
This serves as an evidence that the improved performance of \rsds{} with 
work-stealing is caused by better runtime efficiency and not by better 
schedules.

\setlength{\tabcolsep}{5pt}
\begin{table}
    \caption[caption]{Geometric mean of speedup for experiments  \\
    A (Scheduler) and B (Server); baseline is \dask{}/ws}
    \centering
    \label{tab:geom_mean_speedup}
\begin{tabular}{c|c|r|r|r}
    \textbf{Server} & \textbf{Scheduler} & \textbf{Node count} & \textbf{Worker 
    count} & 
    \textbf{Speedup} \\
    \midrule
    dask & random & 1 & 24 & $0.88\times$ \\
	dask & random & 7 & 168 & $0.95\times$ \\
	rsds & random & 1 & 24 & $1.04\times$ \\
	rsds & random & 7 & 168 & $1.41\times$ \\
	rsds & ws & 1 & 24 & $1.28\times$ \\
	rsds & ws & 7 & 168 & $1.66\times$ \\
    \end{tabular}
\end{table}

\subsection{Scaling}
\begin{figure*}
    \centering
    \includegraphics[width=\textwidth]{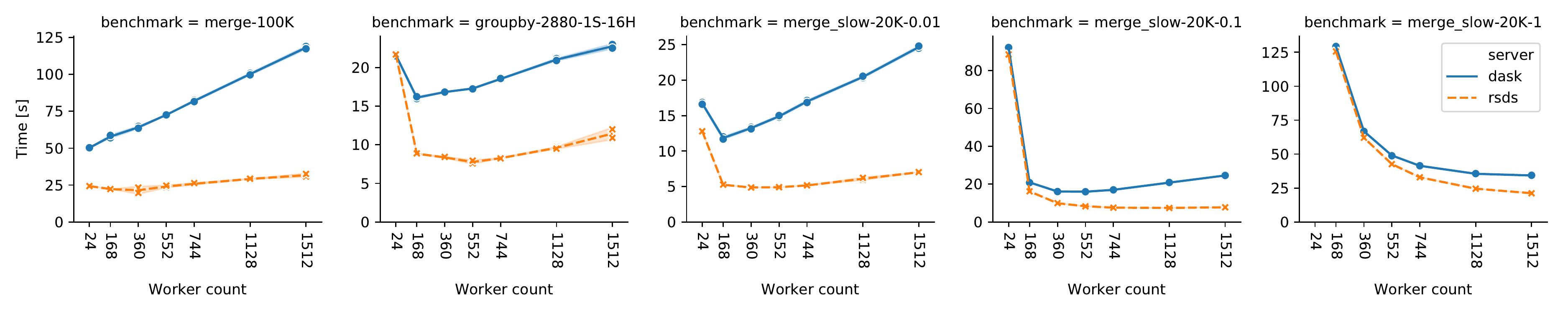}
    \caption{Strong scaling of \dask{} and \rsds{} on selected task graphs}
    \label{fig:scaling}
\end{figure*}

To examine how \dask{} scales to larger clusters and whether the reduced 
overhead of \rsds{} can improve scaling, we have designed an experiment which 
tests the strong scaling of both servers on several cluster sizes ranging from 
1 (24 workers) to 63 nodes (1512 workers). 
The results of  this experiment are shown in Figure~\ref{fig:scaling}. The 
work-stealing algorithm was used for both server implementations.

The first examined task graph is \texttt{merge-100K}, which executes hundred 
thousand trivially short tasks. It is an adversarial case for a scheduler, as 
the tasks are short and  thus the overhead of scheduling and network transfers 
will overcome most parallelism gains. Therefore, increasing the amount of 
workers should not provide a large speedup for this task graph. However, it 
should ideally not slow down the computation to a large extent. We can see that 
\rsds{} scales only up to 15 nodes (360 workers). This is caused by the fact 
that the cost associated with worker management and work-stealing raises with 
an increasing number of workers and 
from some point it starts to dominate because the tasks are too short. \dask{} 
is twice slower when compared to \rsds{} with a single worker node and it is 
four times slower on 63 nodes (1512 workers). Here we can 
see that the inner overhead of \dask{} adds up and its performance is reduced 
significantly with each additional worker node.

Next we examine \texttt{groupby-2880-1S-16H}, which computes an 
analysis of table data using a task graph automatically generated by \dask{} 
by the \texttt{pandas} API. This task graph provides opportunities for 
parallelization, as the individual tasks work on a subset of rows and thus have 
more computational density compared to the \texttt{merge} task graph. However, 
Table \ref{tab:graph_properties} shows that the average computation time is 
still only around 10ms while the average task output is 1 MiB. It 
thus produces considerable network traffic. While both \dask{} and \rsds{} have 
identical performance with a single worker node, \dask{} stops scaling at 7 
nodes and further its performance degrades and eventually 
becomes slower than the single node case. \rsds{} scales up to 23 
worker nodes, hence it is able to utilize three times more workers. With more 
worker nodes the performance of \rsds{} also 
degrades, as the network communication caused by task output 
transfers and work-stealing messages starts to 
dominate the overall execution time.

The third examined task graph is \texttt{merge\_slow-20K}, which executes 
twenty thousand tasks where each task has a fixed duration specified by a 
parameter. Note that 
\texttt{merge} and \texttt{merge\_slow} have the exact same task graph shape. 
The only difference is the duration of each task. We have benchmarked three 
variants of this task graph, with $0.01$, $0.1$ 
and $1$ second tasks. This gives us a better idea of the task granularity 
required 
for \dask{} and \rsds{} to scale effectively. With 10 millisecond tasks, 
\dask{} scales to 7 workers and then its 
performance follows a similar shape as for \texttt{merge-100K}. \rsds{} stops 
scaling at 15 nodes, then its performance drops slightly with more added nodes. 
With 100 millisecond tasks, \rsds{} is able to scale up to 47 worker nodes 
(1128 workers), from that point on its performance stagnates. \dask{} scales 
only up to 23 worker nodes, then the makespan again starts to increase when 
additional workers are added. For the last task graph with one second tasks, 
both \rsds{} and \dask{} scale up to 63 nodes (1512 workers); however, \rsds{} 
is consistently faster on all cluster sizes and its performance in respect to 
\dask{} increases with added worker nodes; it is $1.03\times$ faster than 
\dask{} with 7 nodes and $1.6\times$ faster with 63 nodes.

This shows that for embarrassingly parallel task graphs, \dask{} needs at least 
ca. $100$ millisecond task duration in order to scale up to a larger 
number of workers. Otherwise it will slow down with each added worker node. The 
reduced overhead of \rsds{} enables it to scale better and it also keeps its 
performance relatively stable with an increasing number of workers, even for 
short tasks.

\subsection{Worker overhead isolation}
We have seen that there is considerable overhead in the \dask{} server and that 
lowering that overhead can shorten makespans of executed task graphs. However, 
it is possible that other overheads in the \dask{} runtime might conceal 
additional inefficiency of the server. We test 
both the \dask{} server and \rsds{} with the zero worker -- a minimal worker 
implementation that does not perform any real computation.
Using an idealized worker implementation helps us understand the 
server performance better, because it then remains as the only bottleneck that 
could hinder the execution.

In this experiment, all benchmarks are executed with the zero worker. We use 
only a subset of our benchmark set in this experiment, 
as some of the benchmarks depend on concrete outputs of tasks and thus cannot 
be used with the zero worker, which produces mock data. Note that since 
all of the tasks are computed immediately and have the same output size, the 
only difference between the benchmarks in this experiment is in their task 
graph structure. Also note that the since the tasks are computed immediately, 
any potential attempts to steal a task from a worker will fail. Nonetheless, 
the workers will still respect the initial assignments from the scheduler.

\begin{figure}
	\centering
	\includegraphics[width=0.5\textwidth]{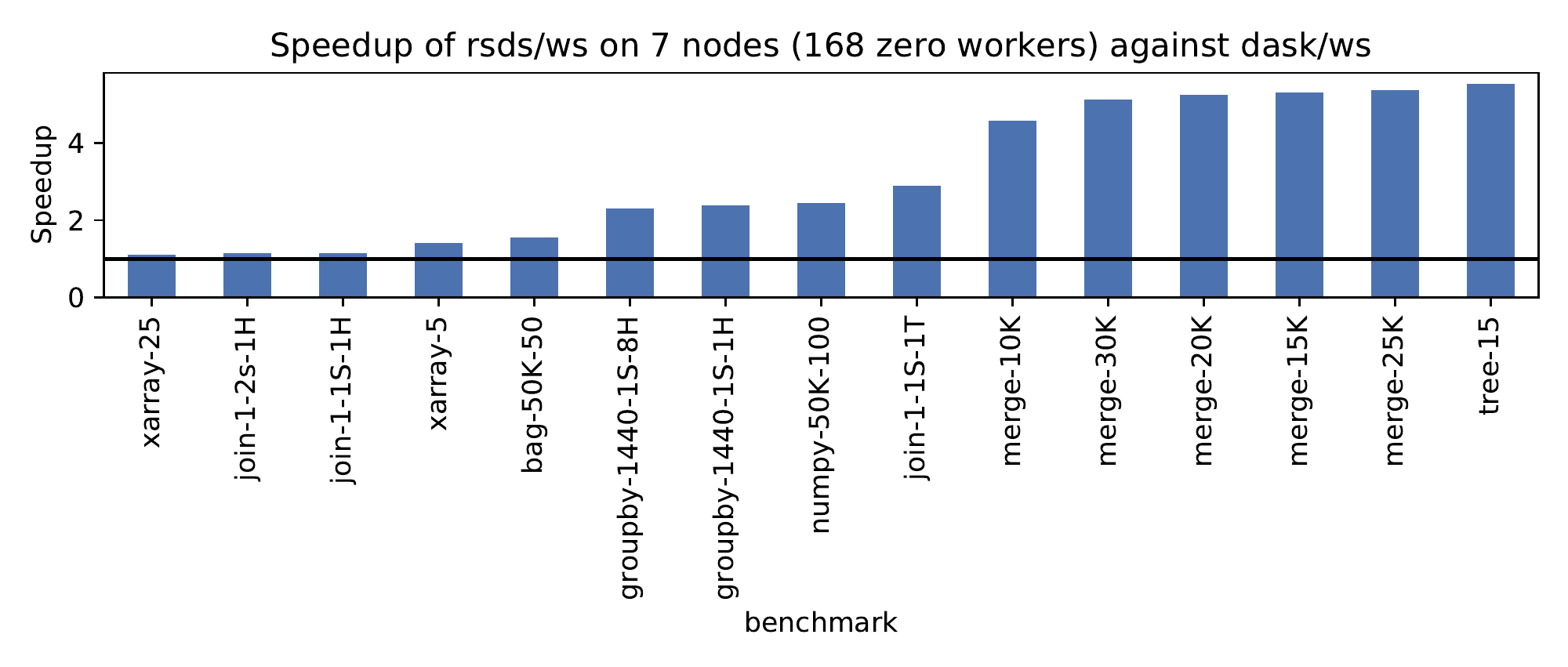}
	\caption{Speedup of \rsds{}/ws over \dask{}/ws when zero worker is used}
	\label{fig:speedup-zw}
\end{figure}

In addition to makespan, we evaluate the average runtime overhead per task 
(AOT), which is calculated as the 
makespan divided by the number of tasks in the task graph. 
Figure~\ref{fig:speedup-zw} shows speedup of 
\rsds{} against \dask{} when the zero worker is used. Results shows that 
\rsds{} is $1.1 - 6\times$ faster. This is a larger difference than when the 
standard worker was used, which suggests that \rsds{} would benefit from a 
faster worker implementation more than the \dask{} server. 

\begin{figure}
	\centering
	\includegraphics[width=0.8\columnwidth]{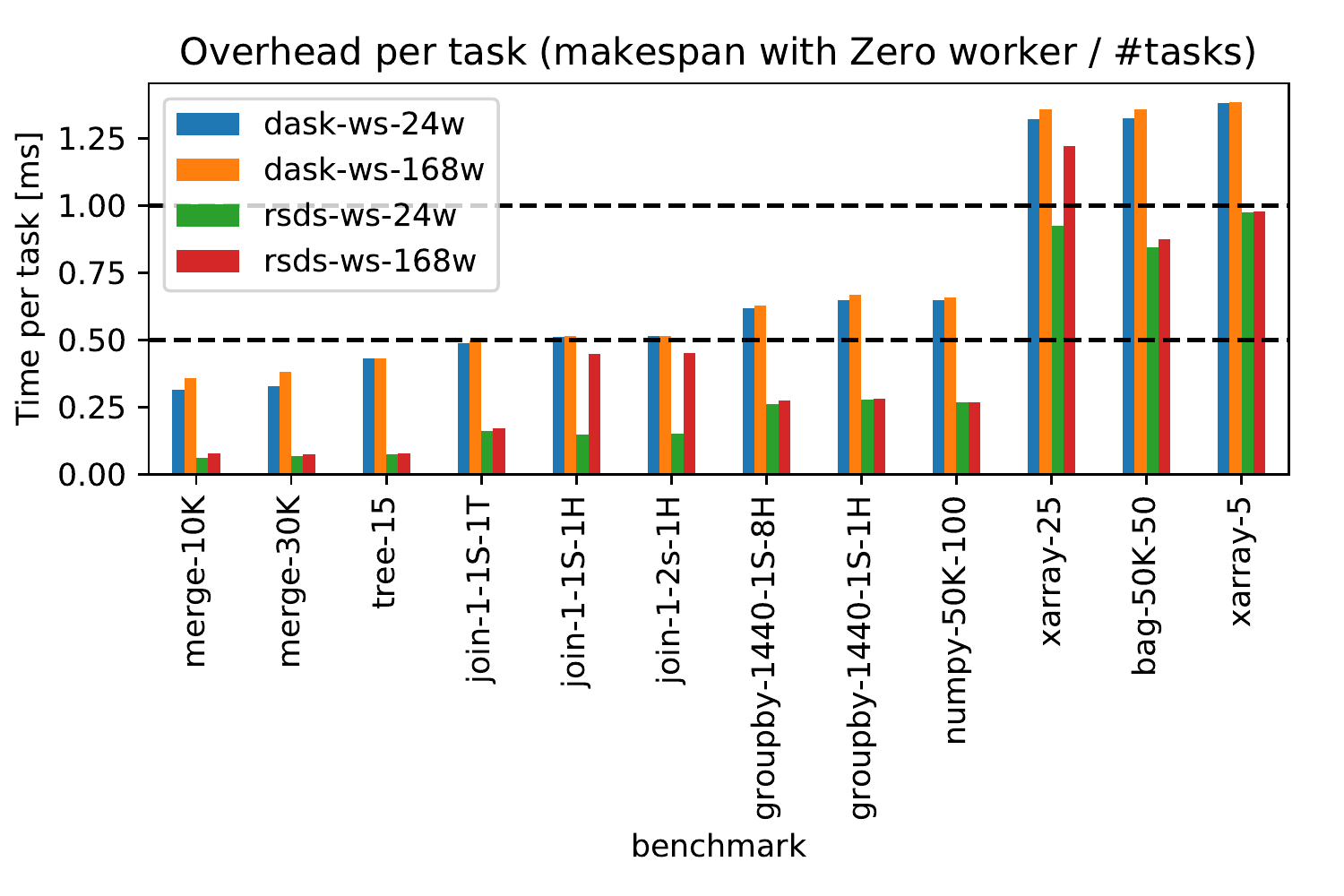}
	\caption{Overhead per task for various cluster sizes and benchmarks}
	\label{fig:tt-all}
\end{figure}

The \dask{} manual states that ``Each task suffers about 1ms of overhead. A 
small computation and a network roundtrip can complete in less than 
10ms.''~\footnote{\url{https://distributed.dask.org/en/latest}}. Our tests 
with AOT show that the overhead is less 
than 1ms for most of our benchmarks. Results are shown in 
Figure~\ref{fig:tt-all}.

\begin{figure}
	\includegraphics[width=0.5\textwidth]{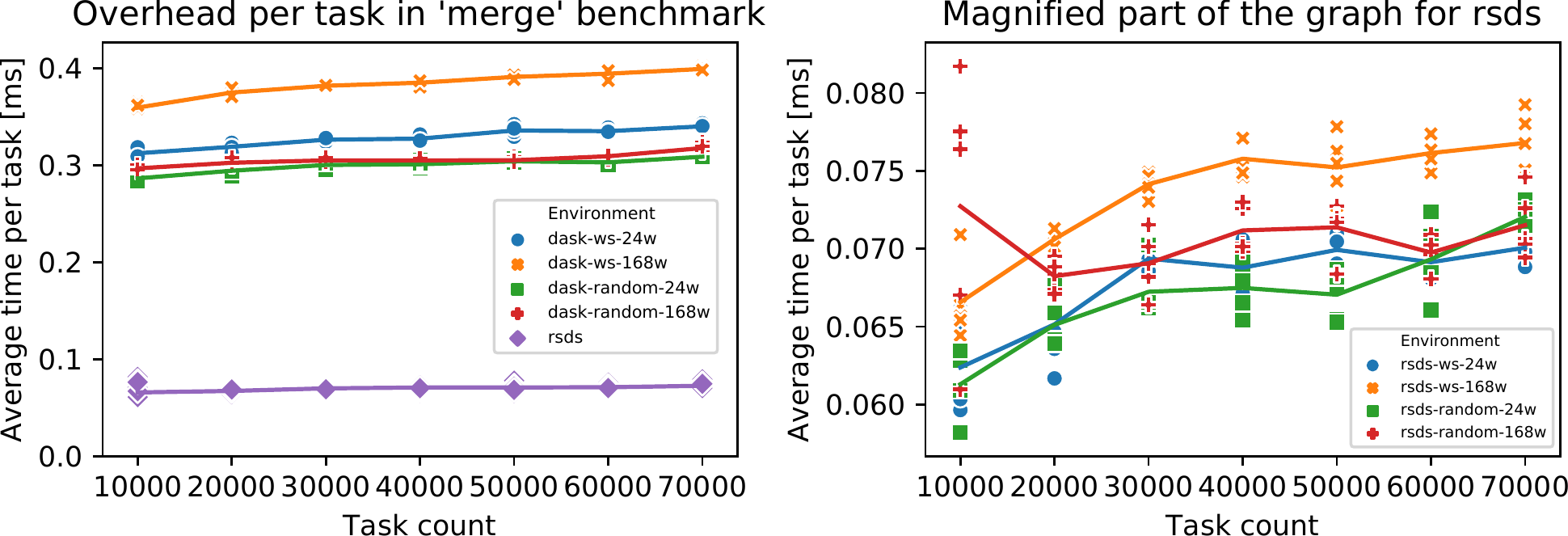}\\

	\includegraphics[width=0.5\textwidth]{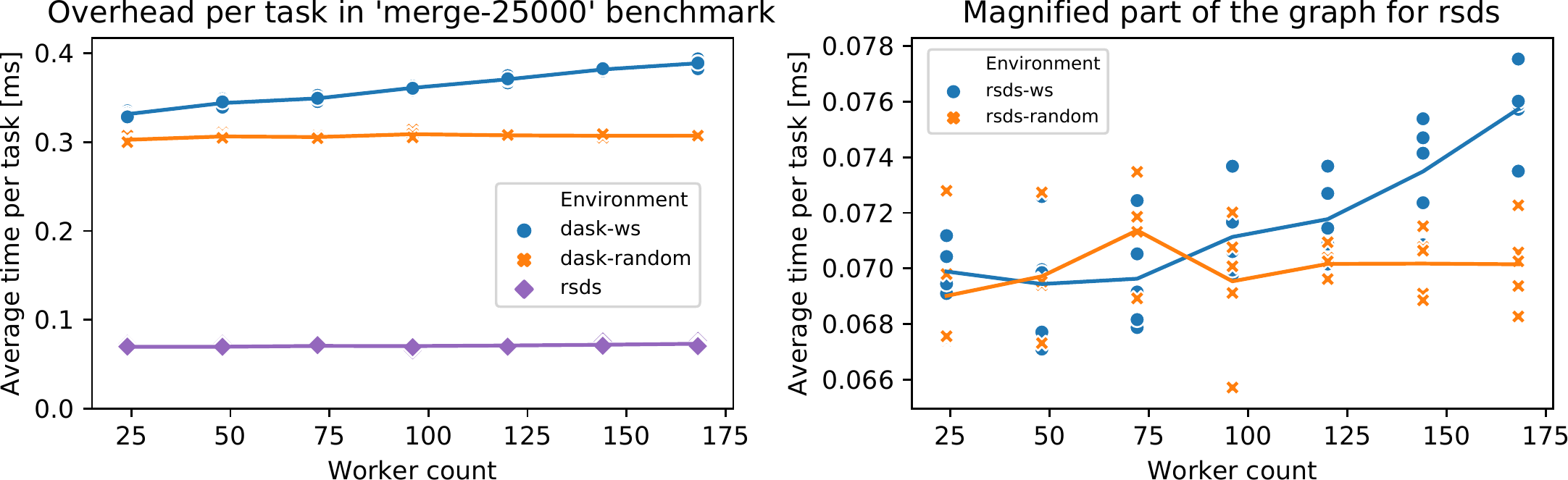}
	\caption{Average task times in the \texttt{merge} benchmark when zero 
		worker is used: (top) scaling task count;
		(bottom) scaling worker count.}
	\label{fig:tt-merge}
\end{figure}

Figure~\ref{fig:tt-merge} shows AOT for the \texttt{merge} benchmark. 
The top graph shows how AOT changes when the number of tasks is increased, the 
bottom graph shows how AOT changes with an increasing number of workers. 

The results show that the random scheduler has less overhead and its 
overhead does not increase as fast with an added number of workers as with the 
work-stealing scheduler, which is expected. This effect is also confirmed by 
the bottom chart of Figure~\ref{fig:tt-merge}, where the overhead of the random 
scheduler stays almost constant when more workers are added.

This data indicates that the general runtime overhead of \dask{} mainly 
grows with an increasing number of tasks, no matter which scheduler is used. On 
the other hand, overhead of the work-stealing scheduler grows primarily with 
the number of workers. In the case of \rsds{}, work-stealing overhead stays 
constant for up to 100 workers, then it also starts to grow with additional 
workers. It also shows an increasing tendency when the number of tasks is 
increased. However, its overhead stays well under \dask{} on all benchmarked 
configurations.

%% file: conclusion.tex
In our paper we have studied the performance of \dask{}, a distributed task 
framework with a central scheduler. We have demonstrated that the benefits of 
a sophisticated scheduler design might be dwarfed by runtime inefficiencies of 
the task framework -- in other words, unless the runtime is properly optimized 
first, it might not be worthwhile to spend too much time on clever scheduler 
designs. We show that for some task graphs even a 
completely random scheduler is competitive with the \dask{} built-in scheduler, 
especially with larger clusters. Our experiment with an idealized worker 
implementation has quantified the per-task overhead of \dask{}. Our experiments 
demonstrate that for an embarrassingly parallel benchmark, the overhead of the 
\dask{} runtime grows with the number of tasks, independently on the used 
scheduler. On the other hand, the performance of the work-stealing scheduler is 
mainly affected by the amount of workers in the cluster.

We have introduced \rsds{}, a drop-in replacement of the \dask{} server 
that outperforms the original server implementation even though it uses a 
simpler and less tuned scheduler. \rsds{} is open-source and can be used to 
accelerate common existing \dask{} workflows without any source code changes.

In future work, we would like to explore the effects of network 
topology and bandwidth on the performance of the \dask{} scheduler. It would 
also be interesting to quantify the effect of improving worker performance on 
the overall workflow runtime.